\documentclass{Pomona2004}

\usepackage{amsmath,amssymb}

\usepackage{rcdcom}
\usepackage[graph]{rcdgraph}

\usepackage[cp1251]{inputenc}
\usepackage[T2A,OT1]{fontenc}
\usepackage[english,russian]{babel}

\def\cal{\mathcal}

\def\currentvolume{?}

\def\ppages{1--10}

\newtheorem{pro}{\sc Proposition}[section]
\newtheorem{rem}{\sc Remark}

\def\d{\delta}

\def\pt#1#2{\frac{\partial #1}{\partial #2}}

\title[Reduction and chaotic behavior of point vortices]
      {Reduction and chaotic behavior of point vortices on a plane and a sphere}
\author[A.\,V.\,Borisov, A.\,A.\,Kilin and I.\,S.\,Mamaev]{}

 \email{borisov@rcd.ru}
  \email{mamaev@rcd.ru}

\begin{document}

\selectlanguage{english}

\maketitle

\centerline{\scshape   A.\,V.\,Borisov, A.\,A.\,Kilin, I.\,S.\,Mamaev}
 \medskip

  {\footnotesize \centerline{ Institute of Computer Science }
  \centerline{ Udmurt State University } \centerline{ 426034
  Izhevsk, Russia } }
 \medskip

 \medskip

\begin{abstract}
We offer a new method of reduction for a system of point vortices on a
plane and a sphere. This method is similar to the classical node
elimination procedure. However, as applied to the vortex dynamics, it requires
substantial modification. Reduction of four vortices on a sphere is given
in more detail. We also use the Poincar\'e surface-of-section technique to
perform the reduction a four-vortex system on a sphere.
\end{abstract}

\section{Equations of motion and first integrals of a vortex system on a
plane} Let us dwell briefly on the basic forms of the equations and first
integrals of the point vortices' motion on a plane. A more complete
account can be found in~\cite{bmk-1,bmk-2}, where the hydrodynamic
assumptions are also given, under which these equations are valid.

The equations of motion of~$n$ point vortices with Cartesian
coordinates~$(x_i,y_i)$ and intensities~$\Gamma_i$ can be written in the
Hamiltonian form:\vspace{-1mm}
\begin{equation}
\label{bmk-eq-1}
\Gamma_i x_i=\pt{\cal H}{y_i},\quad \Gamma_i \dot y_i=-\pt{\cal H}{x_i},\quad 1\le
i\le n,\vspace{-2mm}
\end{equation}
where the Hamiltonian is
\begin{equation}
\label{bmk-eq-2}
{\cal H}=-\frac{1}{4\pi}\sum_{i<j}^n{}\Gamma_i\Gamma_j\ln M_{ij},\q M_{ij}=|\bs r_i-\bs
r_j|^2,\quad\bs r_i=(x_i,y_i).
\end{equation}
Here, the Poisson bracket is\vspace{-3mm}
\begin{equation}
\label{bmk-eq-5}
\{f,g\}=\sum_{i=1}^N\frac{1}{\Gamma_i}\Bigl(\pt{f}{x_i}\pt{g}{y_i}-\pt{f}{y_i}\pt{g}{x_i}\Bigr).
\end{equation}
Since equations~\eqref{bmk-eq-1} are invariant under the group of motions
of plane~$E(2)$, they have, beside the Hamiltonian, three integrals of
motion:\nopagebreak\vspace{-2mm}
\begin{equation}
\label{bmk-eq-3}
Q=\sum_{i=1}^n\Gamma_i x_i,\quad P=\sum_{i=1}^n \Gamma_i y_i,\quad
I=\sum_{i=1}^n \Gamma_i(x_i^2+y_i^2),
\end{equation}
which, however, are not involutive:\nopagebreak\vspace{-2mm}
\begin{equation}
\label{bmk-eq-4}
\{Q,P\}=\sum_{i=1}^N\Gamma_i,\quad \{P,I\}=-2Q,\quad \{Q,I\}=2P.
\vspace{-2mm}
\end{equation}
Hereinafter, it will be more convenient to use instead of~$I$ an integral
of the form\nopagebreak\vspace{-1mm}
\eq[bmk-eq-4a]{
D=\sum_{i<j}^n\Gamma_i\Gamma_j|\bs r_i-\bs
r_j|^2=(\sum_{i=1}^n\Gamma_i)I-P^2-Q^2.
\vspace{-2mm}
}
From these integrals,
one can make two involutive integrals,~$Q^2+P^2$ and~$I$; then, according
to the general theory~\cite{bmk-6}, the system's order can be reduced by
two degrees of freedom. Thus, the three-vortex problem is reduced to a
system with one degree of freedom and, therefore, is integrable (Gr\"obli,
Kirchhoff, Poincar\'e)~\cite{bmk-2}, while the four-vortex problem is
reduced to a system with two degrees of freedom. The latter problem,
generally, is not integrable \cite{bmk-7}.

{\small Effective reduction in the system of four vortices with intensities of the
same sign was done by K.\,M.\,Khanin in~\cite{bmk-4}. He considered two
pairs of vortices, for each of which corresponding action-angle variables
were selected, while the general system was obtained as perturbation of
the two unperturbed problems. Applying this procedure for construction of
the perturbation, he proved (using the methods of KAM-theory) the
existence of quasiperiodic solutions. As a small parameter, he took the
inverse of the distance between the two pairs of vortices.

Reduction in the four-vortex problem in the case when all the four
vortices have equal intensities and in the case when there are two
identical pairs of vortices was done in the papers~\cite{bmk-4}
and~\cite{Aref1}, respectively. A generalization of the latter reduction
to the case of~$N$ vortex pairs is offered in~\cite{Eck1}. In~\cite{Cel},
the KAM\1theory is applied to the reduced equations of the four-vortex
problem.

Reduction by one degree of freedom using the translational invariants~$P$
and~$Q$ was done by Lim in~\cite{bmk-8}. He introduced Jacobi
(barycentric) coordinates (centered, in this case, at the center of
vorticity), which have well-known analogs in the classical~$n$-body
problem in celestial mechanics~\cite{bmk-3}. Note that even this (partial)
reduction made it possible to apply some methods of KAM-theory to
investigation of the motion of point vortices~\cite{bmk-8}.

In the paper~\cite{bmk-9}, a formal Lie algebraic construction was applied to
the case of~$N$ arbitrary vortices to reduce the system's order by two
degrees of freedom.

}\vspace{-2mm}

\section{Reduction on a plane}

Let us reduce the system's order by two degrees of freedom in the problem
of~$N$-vortices of arbitrary intensities. To this end, we will use an
analogy with the planar~$n$-body problem from celestial mechanics.

As is well known, in the~$n$-body problem, one starts with {\it
elimination of the center of mass}~\cite{bmk-3}. To do this, one can
introduce the Jacobi variables
$$
\bs\xi_2=\bs r_2-\bs r_1,\;\bs\xi_3=\bs r_3-\frac{m_1\bs r_1+m_2\bs
r_2}{m_1+m_2},\;\dots,\;\bs\xi_n=\bs r_n-\frac{m_1\bs
r_1+\dots+m_{n-1}\bs r_{n-1}}{m_1+\dots+m_{n-1}},
$$
which define the positions of the bodies in the frame of reference fixed
to the center of mass~$\left(\sum m_i\bs r_i=0\right)$. Clearly, a similar
procedure in the~$n$-vortex problem (masses are replaced with
vorticities~$m_i\to \Gamma_i$) allows us to {\it eliminate the center of
vorticity}. In other words, we use here the group of translations
from~\cite{bmk-8,Limm}. The obtained reduced system is invariant under
rotations (group~$SO(2))$ about the center of vorticity. In celestial
mechanics, the reduction with the rotational symmetry is known as {\it
Jacobi's node elimination} and can be performed in various ways
(see~\mbox{\cite{bmk-3,Uinteker}}).

In vortex dynamics, the above procedure differs substantially (for the
equations of motion are the first order differential equations in time)
and can be performed by using polar coordinates for the Jacobi
variables~($\xi_{ix}=\sqrt{2\rho_i}\cos\vfi_i$,
$\xi_{iy}=\sqrt{2\rho_i}\sin\vfi_i$) with subsequent elimination of the
angle of total rotation of the vortex system as a whole.


\begin{figure}[!ht]
\begin{center}
\cfig<bb=0 0 46.1mm 45.24mm>{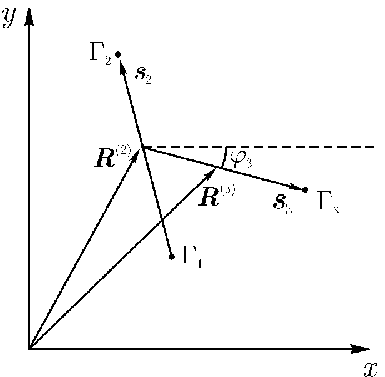}
\caption{}
\label{ris1}
\end{center}
\vspace{-6mm}
\end{figure}

\begin{pro}
\label{pred1} A planar system of~$N$ arbitrary vortices of nonzero total
intensity~($\sum_{i=1}^{N}\Gamma_i\ne0$) allows reduction by two degrees
of freedom. The canonical variables of the reduced system are
\begin{equation}
\label{new-01}
\begin{aligned}
q_{i} & =\rho_{i+2}\frac{\Gamma_{i+2}\sum_{k=1}^{i+1}\Gamma_k}{\sum_{k=1}^{i+2}\Gamma_k},\quad
\psi_{i}=\vfi_{i+2}-\vfi_2,\quad i=1,\ldots, N-2,
\end{aligned}\vspace{-3mm}
\end{equation}
where\vspace{-3mm}
\begin{equation}
\label{new-02}
\rho_i=\frac{1}{2}|\br_i-\bR^{(i-1)}|^2,\q
\vfi_i=\arctg\left(\frac{y_i-R^{(i-1)}_y}{x_i-R^{(i-1)}_x}\right),\q
i=2\ldots N,
\end{equation}
Here, $\bR^{(i)}=\frac{\sum_{j=1}^i\Gamma _j\br_j}{\sum_{j=1}^i\Gamma_j}$
defines the center of vorticity of the~$i$ vortices, while~$\br_i$ is the
radius-vector of the~$i$th vortex (see Fig.~$\ref{ris1}$).
\end{pro}
\proof

Let us first calculate the Poisson bracket of~$\rho_i$ and~$\vfi_i$:
\begin{equation}
\label{new-04}
\{\rho_i,\vfi_j\}=\delta_{ij}\frac{\sum_{k=1}^i\Gamma_k}{\Gamma_i\sum_{k=1}^{i-1}\Gamma_k},\q
i=2,\,\ldots,\,N.
\end{equation}
If the numerator and the denominator do not vanish, we normalize the
variables~$\rho_i$ so that the~$q_i,\,\psi_i$ found from~\eqref{new-01}
are canonical. It is easy to show that if~$\sum_{i=1}^{N}\Gamma_i\ne0$,
then the vortices can be divided into a few groups, such that for each
group~${\suml_{k=1}^{i}\Gamma_k\ne0}$, $i=1,\,\ldots,\,N-1$.

Now we show that on a commonn level surface of the integrals~$Q$, $P$,
and~$D$~\eqref{bmk-eq-3},~\eqref{bmk-eq-4a}, the
Hamiltonian~\eqref{bmk-eq-2} can be expressed in terms of the
variables~\eqref{new-01}. Indeed, the squared mutual distances~$M_{ij}$
can be expressed in terms of the vectors~$\bs s_i,\,i=1\ldots N$, from
the~$i$th vortex to the center of vorticity of the~$(i-1)$th vortex'
subsystem:\vspace{-3mm}
\begin{equation}
\label{new-003}
M_{ij}=\left|\bs s_i-\bs s_j+
\sum_{k=j}^{i-1}\frac{\Gamma_k\bs s_k}
{\sum_{l=1}^k\Gamma_l}\right|^2,\quad i>j,\quad
\bs s_i=\br_i-\frac{\sum_{j=1}^{i-1}\Gamma
_j\br_j}{\sum_{j=1}^{i-1}\Gamma_j}.
\end{equation}
According to~\eqref{new-02}, $\bs
s_i=(\sqrt{2\rho_i}\cos\vfi_i,\,\sqrt{2\rho_i}\sin\vfi_i)$; here,~$\rho_i$
and~$\vfi_i$ are expressed in terms of~$q_i$,~$\psi_i$ and the
angle~$\psi_0=\vfi_2$ of total rotation of the vortex system as a whole
about the common center of vorticity:
\begin{equation}
\label{new-005}
\begin{gathered}
\rho_2=
\left(\frac{D}{2\sum_{i=1}^{N}\Gamma_i}-\sum_{k=1}^{N-2}q_k\right)\frac{\Gamma_1+\Gamma_2}
{\Gamma_1\Gamma_2},\quad \rho_i =
\frac{\sum_{k=1}^i\Gamma_k}{\Gamma_i\sum_{k=1}^{i-1}\Gamma_k}q_{i-2},\\
\vfi_i=\psi_{i-2}+\psi_0,\quad i=3,\ldots, N.
\end{gathered}
\end{equation}
Since the squared mutual distances are expressed in terms of all possible
scalar products,~$(\bs s_i,\,\bs
s_j)=2\sqrt{\rho_i\rho_j}\cos(\vfi_i-\vfi_j)$, they do not depend
on~$\psi_0$. Thus, on the common level surface of the system's integrals,
the squared mutual distances and, consequently, the system's Hamiltonian
are expressed in terms of the variables~\eqref{new-01}.
\qed

Let us discuss in more detail the special
case~$\sum\limits_{i=1}^N\Gam_i=0$, which has no analog in celestial
mechanics (since bodies' masses are always positive,~$m_i>0)$. In this
case we say that the center of vorticity is at infinity, and the
variables~\eqref{new-01} are undefined (since one of the denominators
vanishes). A corresponding reduction is described in the following way:

\begin{pro}
When~$\sum_{j=1}^{N}\Gamma_j=0$, the canonical variables of the reduced
system (those obtained from reduction by two degrees of freedom),
$\widetilde{\rho}_i,\,\widetilde{\vfi}_i,\, i=1,\,\ldots,\,N-2$, are given
by
\begin{equation}
\label{new-03plus}
\widetilde{\rho}_i=\frac{\Gamma_i\sum_{k=1}^{i-1}\Gamma_k}{\sum_{k=1}^i\Gamma_k}\rho_{i+1},\q
\widetilde{\vfi}_i=\vfi_{i+1},\q i=1,\,\ldots,\,N-2,
\end{equation}
where $\rho_i$ and $\vfi_i$ are given by~\eqref{new-02}.
\end{pro}

\proof As we showed earlier, the Poisson bracket of~$\rho_i$
and~$\vfi_i$~\eqref{new-02} is defined by~\eqref{new-04}.
Now we have~$\{\rho_N,\,\vfi_N\}=0$, while~$\rho_N$ and~$\vfi_N$ are the
integrals of motion. They are related to the standard integrals of
motion~\eqref{bmk-eq-3} and~\eqref{bmk-eq-4a} in the following way:
\begin{equation}
\label{new-07}
P=\Gamma_N\sqrt{2\rho_N}\cos\vfi_N,\,
Q=\Gamma_N\sqrt{2\rho_N}\sin\vfi_N,\,
D=-\Gamma_N^2\rho_N.
\end{equation}

As in the previous Proposition, the squared mutual distances~$M_{ij}$ are
expressed in terms of~$\bs
s_i=(\sqrt{2\rho_i}\cos\vfi_i,\,\sqrt{2\rho_i}\sin\vfi_i)$, according
to~\eqref{new-003}. Hence, the system's Hamiltonian on a common level
surface of the system's integrals depends on ${2(N-2)}$
variables,~($\rho_i,\,\vfi_i,\,i=2\ldots N-1$), and on the values of the
two integrals,~$\rho_N$ and $\vfi_N$. \qed

There is an even more special (but no less interesting) case:
$\sum_{j=1}^{N}\Gamma_j=0$ and $D=0$. Since in this case we have three
involutive integrals~\eqref{bmk-eq-3}, we can reduce the system's order by
three degrees of freedom. For example, the four-vortex problem yields
under these conditions a special case of integrability~\cite{82,Eck},
while a system of five point vortices is reduced to a system with two
degrees of freedom. Indeed,
\begin{pro}
When $\sum_{j=1}^{N}\Gamma_j=0$ and~$D=0$, a system allows reduction by
three degrees of freedom. The canonical variables of the reduced system
are defined by~\eqref{new-01},~$i=1\ldots N-3$.
\end{pro}
\proof As we showed in the previous Proposition, in the case
of~$\sum_{j=1}^{N}\Gamma_j=0$, the change of variables~\eqref{new-02}
reduces the system's dimensionality by two.
Upon the reduction procedure, we obtain the
variables~$\rho_k,\,\vfi_k,\,k=2\ldots N-1$. Now, to reduce the system's
order by one more degree of freedom, we should the
variables~\eqref{new-01} defined now for~$i=1\ldots N-3$. Using the
methods of Proposition~\ref{pred1}, we can show that, on the common level
surface of the first integrals, the Hamiltonian is expressed in terms of
the variables~$q_i,\,\psi_i,\,i=1\ldots N-3$, and depends on the
parameters~$\rho_N$ and~$\vfi_N$. \qed

\section{The equations of motion and first integrals of a vortex system
on~$\mathbb{S}^2$}

For $n$ point vortices on~$\mathbb{S}^2$, the Hamiltonian equations of
motion in terms of spherical coordinates~$(\theta_i,\varphi_i)$ can be
written as follows~\cite{Bog2}:
\begin{equation}
\label{bmk-eq-6} \dot{\theta}_i=\{\theta_i,{\cal H}\},\quad
\dot{\varphi}_i=\{\varphi_i,{\cal H}\},\quad i=1,\ldots,n,
\end{equation}
with the Poisson bracket
\begin{equation}
\label{bmk-eq-7}
\{\varphi_i,\cos\theta_k\}=\frac{\delta_{ik}}{R^2\Gamma_i},
\end{equation}
where $\Gamma_i$ are the vortices' intensities, and the Hamiltonian is
\begin{equation}
\label{bmk-eq-8} {\cal
H}=-\frac1{4\pi}\sum_{i<k}^n\Gamma_i\Gamma_k\ln\left(M_{ik}\right)=-\frac1{4\pi}\sum_{i<k}^n\Gamma_i\Gamma_k\ln\left(4R^2\sin^2\frac{\gamma_{ik}}2\right).
\end{equation}
Here, $R$ is the radius of the sphere, $M_{ij}$ is the squared distance
between the~$i$th and the~$j$th vortices, measured along the chord,
and~$\gamma_{ik}$ is the angle between the vectors from the sphere's
center to the point vortices~$i$ and~$k$,
$$
\cos\gamma_{ik}=\cos\theta_i\cos\theta_k+\sin\theta_i\sin\theta_k\cos(\varphi_i-\varphi_k).
$$
Beside the Hamiltonian, the equations~(\ref{bmk-eq-6}) admit three
additional independent involutive integrals:
\begin{equation}
\label{bmk-eq-9}
F_1=R\sum_{i=1}^n{}\Gamma_i\sin\theta_i\cos\varphi_i,\q
F_2=R\sum_{i=1}^n{}\Gamma_i\sin\theta_i\sin\varphi_i,\q
F_3=R\sum_{i=1}^n{}\Gamma_i\cos\theta_i.
\end{equation}
The vector~$\bs F$, with components~$(F_1,\,F_2,\,F_3)=\bs F=\sum\Gamma_i
\bs r_i$ (where~$\bs r_i$ are the radius-vectors of the vortices), is
called the \emph{moment of vorticity\/}, its components commuting in the
following way:
\begin{equation}
\label{bmk-eq-10} \{F_i,F_j\}=\frac1{R}\varepsilon_{ijk}F_k.
\end{equation}
As in the planar case, the system can be reduced by two degrees of
freedom, using involutive integrals, eg.,~$F_3$ and~$\bs F^2$.

Thus, for the case of three vortices, we obtain a completely integrable
system (this system was independently found and studied
in~\cite{Bor_Leb1,Bor_Leb2,95}). The four-vortex problem is reduced to a
system with two degrees of freedom and, generally, is not integrable~\cite{Bagrets}.\vspace{-3mm}

\section{Reduction on a sphere}

On a sphere, as distinct from a plane, it is impossible to regard symmetry
transformations as translations and rotations (there are only rotations);
nevertheless, the above reduction algorithm (which generalizes the Jacobi
reduction) can itself be generalized. As above, we will consider in
succession the moment of vorticity of two, three,~\dots, and~$n$ vortices:
$$
\bs F_2=\Gam_1\bs r_1+\Gam_2\bs r_2,\q\dots,\q\bs F_n=\Gam_1\bs
r_1+\dots+\Gam_n\bs r_n=\bs F,
$$
where $\br_i\in\mathbb R^3$ are the Cartesian coordinates of the vortices
on a sphere embedded in~$\mathbb R^3$. The squared moments~$\bs F_k^2$,
$k=2\dts n$, have the following (obvious) properties:

$1^\circ.$ all $\bs F_k^2$ commute with each other\vspace{-2mm}
$$
\{\bs F_k^2,\,\bs F_m^2\}=0;
$$

$2^\circ.$ a squared moment $\bs F_k^2$ commutes with the coordinates of
all the vortices numbered~$1,2\dts k$
$$
\{\bs F_k^2,x_i\}=\{\bs F_k^2,y_i\}=\{\bs F_k^2,z_i\}=0,\q i=1\dts k.
$$

Thus, the squared moments~$\bs F_2^2\dts\bs F_{n-1}^2$ are invariant under
the action of the group~$SO(3)$, commute with each other, and their number
is half the dimension of the reduced system. Using~$2^\circ$, it is easy
to add some relative angular variables of the reduced system to this set.

\vspace{-2mm}

\fig<bb=0 0 43.3mm 38.0mm>{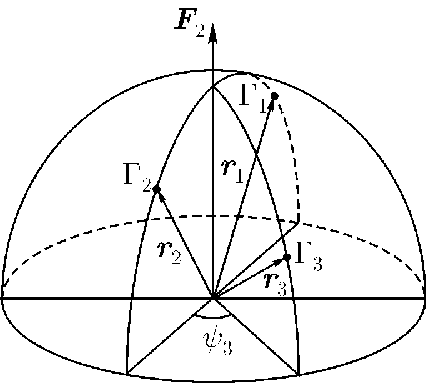}[\label{pic2}]

\vspace{-7mm}

\begin{pro}
A system of $N$ vortices on a sphere,
when~$\bF_{N}=\sum_{i=1}^{N}\Gamma_i\br_i\ne0$, allows reduction by two
degrees of freedom, using the canonical variables~$\rho_i,\,\psi_i$ given
by\vspace{-3mm}
\begin{equation}
\label{new2-01}
\begin{gathered}
\rho_i=|\bF_{i+1}|,\q \tg\psi_i = \frac{\rho_{i}({\bs F}_{i+1},\br_{i+1}\times \br_{i+2})}{(\br_{i+1}\times
{\bs F}_{i+1},\br_{i+2}\times {\bs F}_{i+1})},\q i=1,\,\ldots,\,N-2,
\end{gathered}
\end{equation}
where
$\psi_i$ is the angle between the planes~$(\bF_{i+1},\br_{i+2})$
and~$(\bF_{i+1},\br_{i+1})$ (see Fig. \ref{pic2}).
\end{pro}
\proof 1. It can be shown straightforwardly from~\eqref{new2-01} that the
variables~$\rho_i,\,\psi_i$ commute in the following way:\nopagebreak\vspace{-2mm}
\begin{equation}
\label{k-2-02b}
\{\rho_i,\rho_j\}=\{\psi_i,\psi_j\}=0,\quad \{\rho_i,\psi_j\}=\frac{\delta_{ij}}{R}.
\end{equation}

2. Then we show that on the common level surface of the integrals~$\bF_N$,
the Hamiltonian~\eqref{bmk-eq-8} can be expressed in terms of the
variables~\eqref{new2-01}. Since~$\br_i$ and~$\bF_i$, $i=2,\,\ldots,\,N$,
are linearly related:\vspace{-2mm}
$$
\br_i=\frac{1}{\Gamma_i}(\bF_i-\bF_{i-1}),\q i=2,\,\ldots,\,N,
$$
the squared mutual intervortical distances~$M_{ij}$ can be expressed in
terms of scalar products
of~$\bF_1=\Gamma_1\br_1,\,\bF_2,\,\ldots,\,\bF_N$. So, to prove the
Proposition, it is sufficient to show that these scalar products on the
common level surface of the integral~$\bF_N$ are completely defined by the
variables~\eqref{new2-01}.

Consider an algorithm for construction of the vectors~$\bF_i,\,i=1\ldots
N-1$, using the known variables~\eqref{new2-01} and the values of~$\bF_N$.
By construction, the angle~$\psi_{i-2}$ is the angle between the
planes~$(\bF_{i-1},\br_i)$ and~$(\bF_{i-1},\br_{i-1})$, or, what is the
same, between the planes~$(\bF_{i-1},\bF_i)$ and~$(\bF_{i-1},\bF_{i-2})$.
Hence,\nopagebreak\vspace{-2mm}
\begin{gather}
\bF_{i-2}=\bF_{i-1}\frac{(\bF_{i-1},\,\bF_{i-2})}{\bF_{i-1}^2}+
\Bigl({\bs n}_1\sin\psi_{i-2}+{\bs n}_2\cos\psi_{i-2}\Bigr)\sqrt{\bF_{i-2}^2-\frac{(\bF_{i-1},
\,\bF_{i-2})^2}{\bF_{i-1}^2}},\notag\\
\label{new2-02}
{\bs n}_1=\frac{\bF_{i}\times\bF_{i-1}}{|\bF_{i}\times\bF_{i-1}|},\q
{\bs n}_2=\frac{\bF_{i-1}\times(\bF_{i}\times\bF_{i-1})}{|\bF_{i-1}
\times(\bF_{i}\times\bF_{i-1})|}.
\end{gather}
Using the definitions of $\bF_i$ and $\rho_i$ \eqref{new2-01}, we find
that\nopagebreak\vspace{-2mm}
\begin{equation}
\label{new2-03}
(\bF_{i},\bF_{i-1})=\frac12(\rho_{i-1}^2+\rho_{i-2}^2-\Gamma^2_{i}R^2),\q
\bF_{i}^2=\rho_{i-1}^2.\vspace{-1mm}
\end{equation}
Substituting these relations into~\eqref{new2-02}, we obtain the recurrent
expression for~$\bF_{i-2}$ in terms of~$\bF_{i-1}$, $\bF_{i}$, $\rho_i$,
and~$\psi_i$:
\begin{equation}
\label{new2-04}
\begin{aligned}
\bF_{i-2}&=\bF_{i-1}\frac{\rho_{i-2}^2+\rho_{i-3}^2-\Gamma_{i-1}^2R^2}{2\rho_{i-2}^2}+
\Bigl({\bs n}_1\sin\psi_{i-2}+{\bs n}_2\cos\psi_{i-2}\Bigr)\times\\
&\times\sqrt{\rho_{i-3}^2-\frac{(\rho_{i-2}^2+\rho_{i-3}^2-\Gamma_{i-1}^2R^2)^2}{4\rho_{i-2}^2}},\q
i=3,\,\ldots,\,N,
\end{aligned}
\end{equation}
where $\rho_0=|\bF_1|=\Gamma_1R$.

With~\eqref{new2-04} it is easy to show that\vspace{-1mm}
\begin{equation}
\label{new2-04a}
\bF_{i}=\alpha_i\bF_{N}+\beta_i\bF_{N-1}+\gamma_i\bF_{N}\times\bF_{N-1},\q i=1,\,\ldots,\,N-2,
\end{equation}
where the coefficient~$\alpha_i,\,\beta_i,\,\gamma_i$ are expressed in
terms of the coordinates~\eqref{new2-01}. Hence, all the scalar
products~$(\bF_i,\,\bF_j)$ are expressed in terms of the
coordinates~\eqref{new2-01} and the values of~$(\bF_N,\,\bF_N)$,
$(\bF_{N-1},\,\bF_{N-1})$, and~$(\bF_N,\,\bF_{N-1})$.
Using~\eqref{new2-03}, we can express these values in terms of the
coordinates~\eqref{new2-01} and the constants of the integrals~$\bF_N$. In
this way, the scalar products~$(\bF_i,\,\bF_j)$ (and, consequently, the
mutual intervortical distances) would depend only on the
variables~\eqref{new2-01}. Hence, the variables~\eqref{new2-01} allow the
reduction by two degrees of freedom. \qed

Let us now discuss the special case of further reduction by one more
degree of freedom. This reduction is quite similar to the case of planar
motion of the vortices with~$\sum_{j=1}^{N}\Gamma_j=0$, $D_N=0$. We have
\begin{pro}
When~$\bF_{N}=\sum_{i=1}^{N}\Gamma_i\br_i=0$, the system~\eqref{bmk-eq-6}
allows reduction by three degrees of freedom. The canonical variables of
the reduced system are given by~\eqref{new2-01} with~$i=1,\,\ldots,\,N-3$.
\end{pro}
\proof To prove the Proposition, we show, as in the previous case, that
the scalar products~$(\bF_i,\,\bF_j)$ depend only on the
variables~$\rho_i,\,\psi_i,\,i=1,\,\ldots,\,N-3$.

Using~\eqref{new2-04}, the vectors~$\bF_i$ can be expressed in terms
of~$\bF_{N-1}$ and~$\bF_{N-2}$:
\begin{equation}
\label{new2-05}
\bF_{i}=\widetilde\alpha_i\bF_{N-1}+\widetilde\beta_i\bF_{N-2}+
\widetilde\gamma_i\bF_{N-1}\times\bF_{N-2},\q i=1,\,\ldots,\,N-3,
\end{equation}
where $\widetilde\alpha_i,\,\widetilde\beta_i$, and $\widetilde\gamma_i$
depend on the coordinates~$\rho_i,\,\psi_i,\,i=1,\,\ldots,\,N-3$. Thus,
the scalar products~$(\bF_i,\,\bF_j)$ are expressed in terms of the
coordinates~$\rho_i,\,\psi_i,\,i=1,\,\ldots,\,N-3$,
and~$(\bF_{N-1},\,\bF_{N-1})$, $(\bF_{N-2},\,\bF_{N-2})$,
and~$(\bF_{N-1},\,\bF_{N-2})$. Here, $\bF_{N-1}=-\Gamma_N\br_N$,
therefore,~$\rho_{N-2}=|\bF_{N-1}|=R|\Gamma_N|$ is an integral of motion.
Hence, using~\eqref{new2-03}, we can express the scalar
products~$(\bF_{N-1},\,\bF_{N-1})$, $(\bF_{N-2},\,\bF_{N-2})$, and
$(\bF_{N-1},\,\bF_{N-2})$ in terms
of~$\rho_i,\,\psi_i,\,i=1,\,\ldots,\,N-3$. In this way, the mutual
intervortical distances depend only
on~$\rho_i,\,\psi_i,\,i=1,\,\ldots,\,N-3$, and the
transformation~\eqref{new2-01} allows the reduction by three degrees of
freedom. \qed

Note that under the specified conditions, a four-vortex system on a sphere
is
integrable~\cite{bmk-2,bmk-10}.\vspace{-2mm}

\section{Explicit reduction of a four-vortex system on a
sphere}

Now let us present a reduced system of four vortices on a sphere in
explicit form. The system's Hamiltonian is expressed in terms of the
mutual distances by~\eqref{bmk-eq-8}. The mutual distances between four
vortices on a sphere, given in terms of the canonical variables of the
reduced system~($\{\rho_i,\psi_j\}=R^{-1}\delta_{ij}$, $i,j=1,2$) and the
squared integral~$\bs F^2=\bs F_4^2=c^2=\const$ read:
\begin{gather}
M_{12} = \frac{(\Gamma_1+\Gamma_2)^2R^2-\rho_1^2}{\Gamma_1\Gamma_2},\quad
M_{23}=\frac{((\Gamma_2+\Gamma_3)^2-\Gamma_1^2)R^2-\rho_2^2+2(\bs F_1\bs F_3)}
{\Gamma_2\Gamma_3},\notag\\
M_{13}=\frac{((\Gamma_1{+}\Gamma_3)^2{-}\Gamma_2^2{-}\Gamma_3^2)R^2
{+}\rho_1^2-2(\bs F_1,\bs F_3)}
{\Gamma_1\Gamma_3},\,\,
M_{14}=2R^2{+}2\frac{(\bs F_1,\bs F_3){-}(\bs F_1,\bs
F)}{\Gamma_1\Gamma_4},\notag\\
M_{34}=\frac{(\Gamma_3+\Gamma_4)^2R^2-\rho_1^2-c^2+
2(\bs F_2,\bs F)}{\Gamma_3\Gamma_4},\\
M_{24}=\frac{(2\Gamma_2\Gamma_4-\Gamma_3^2)R^2 +\rho_1^2+\rho_2^2-2(\bs
F_2,\bs F)+2(\bs F_1,\bs F)-2(\bs F_1,\bs F_3)}{\Gamma_2\Gamma_4}.
\notag
\end{gather}
Here, the vortices' intensities are arbitrary. The scalar products
are:\nopagebreak\vspace{-1mm}
\begin{gather}
(\bs F_1,\bs F_2)=\frac{1}{2}(\rho_1^2+\Gamma_1^2R^2-\Gamma_2^2R^2),\quad
(\bs F_2,\bs F_3)=\frac{1}{2}(\rho_2^2+\rho_1^2-\Gamma_3^2R^2),\notag\\
(\bs F_3,\bs F)=\frac{1}{2}(c^2+\rho_2^2-\Gamma_4^2R^2),\notag\\
(\bs F_2,\bs F)=\frac{(\bF_3,\bF)(\bF_2,\bF_3)}{\rho_2^2}+\Bigl(
c^2-\frac{(\bF_3,\bF)^2}{\rho_2^2}\Bigr)\cos\psi_2\sqrt{\frac{\rho_1^2\rho_2^2
-(\bF_2,\bF_3)^2}{\rho_2^2c^2-(\bF_3,\bF)^2}},\notag\\
(\bs F_1,\bs F_3)=\frac{(\bF_2,\bF_3)(\bF_1,\bF_2)}{\rho_1^2}+\Bigl(
\rho_2^2-\frac{(\bF_2,\bF_3)^2}{\rho_1^2}\Bigr)\cos\psi_1\sqrt{\frac{\rho_1^2
\Gamma_1^2R^2 -(\bF_1,\bF_2)^2}{\rho_1^2\rho_2^2-(\bF_2,\bF_3)^2}},\notag\\
(\bs F_1,\bs F)=\frac{(\bF_2,\bF)(\bF_1,\bF_2)}{\rho_1^2}+\Bigl(
(\bF_3,\bF)-\frac{(\bF_2,\bF)(\bF_2,\bF_3)}{\rho_1^2}\Bigr)
\cos\psi_1\times\\
\times\sqrt{\frac{\rho_1^2\Gamma_1^2R^2
-(\bF_1,\bF_2)^2}{\rho_1^2\rho_2^2-(\bF_2,\bF_3)^2}}
+\frac{\sin\psi_1\sin\psi_2}{\rho_1\rho_2}
\sqrt{(\rho_1^2\Gamma_1^2R^2-(\bF_1,\bF_2)^2)(\rho_2^2 c^2-(\bF_3,\bF)^2)}.
\notag
\end{gather}

\fig<bb=0 0 105.8mm 175.3mm>{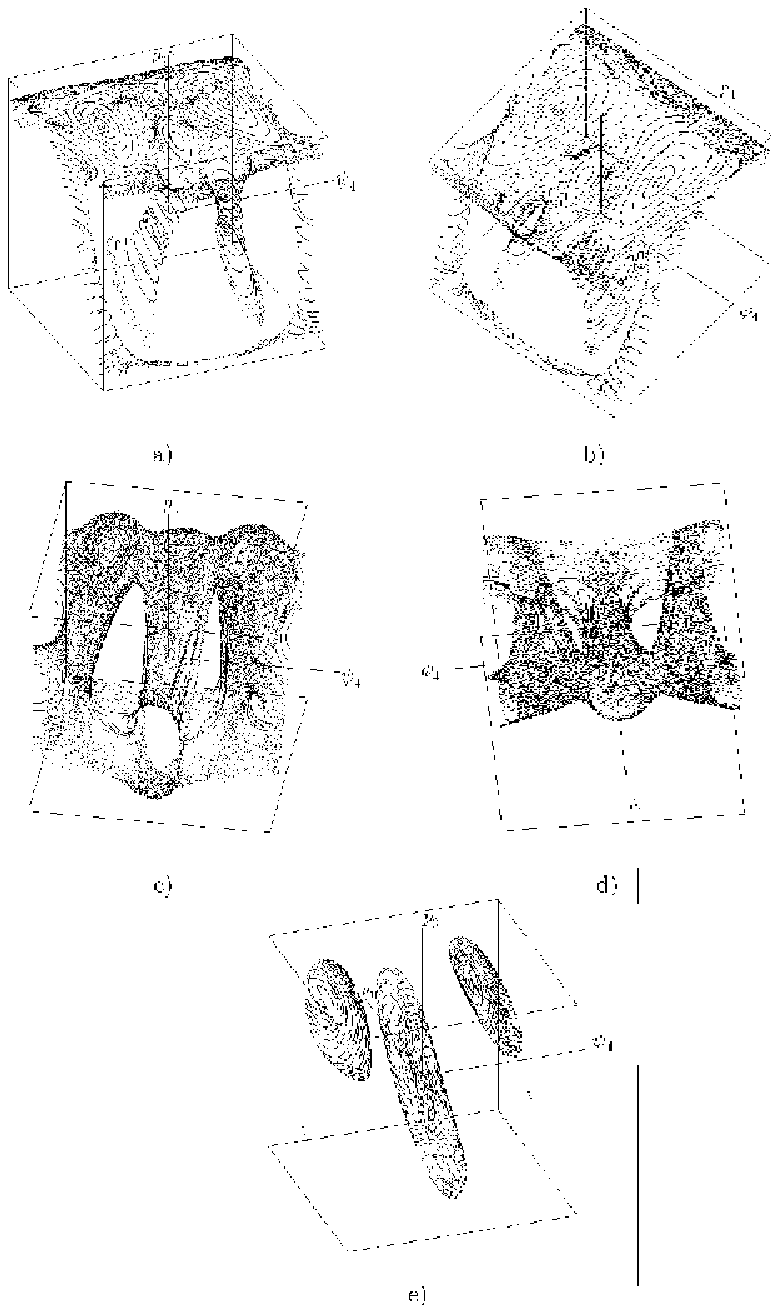}[The Poincar\'e maps for the problem
of four vortices of equal intensities on a sphere,~$D=3.55$. The plane of
section is~$\psi_1=\frac{\pi}{2}$. The energy values are:~$E=0.8$ (Figs. a
and b),~$E=0.67$ (Figs. c and d),~$E=0.64$ (Fig. e).\vspace{-3mm}]

\section{The Poincar\'e section of a four-vortex system on a sphere}

The obtained systems of reduced canonical variables can be applied to
various analytical and numerical studies. Consider, for example, chaos in
the system of four vortices on a sphere.

Below, we present Poincar\'e surface-of-section plots for a reduced system
of four vortices on a sphere. As far as we know, such maps have not been
plotted before.
As a plane of section, we choose~$\psi_1=\const$. The intersection between
this plane and the isoenergetic surface~${\cal
H}(\rho_1,\psi_1,\rho_2,\psi_2)=E=\const$ is some two-dimensional surface
(generally, disconnected) in the space~$\rho_1,\rho_2,\psi_2$. The
system's phase flow on this surface generates a Poincar\'e map. As a rule,
the surface is complicated, therefore, we give here a 3D view of the
Poincar\'e map without projecting it on a plane.

\begin{rem}
Construction of a Poincar\'e section in three-dimensional space allows us
to ignore various singularities of the projection and does not result in
emergence of fictitious objects, mentioned in~\cite{Aref1}~(namely,
crescent-shaped tori, which are due to the projection's singularities).
\end{rem}

Figure~\ref{pic4.eps} shows the phase portraits in the
space~$\rho_1,\rho_2,\psi_2$ for the case of four equal vortices
and~$D=3.55$. The plane of section is~$\psi_1=\frac{\pi}{2}$. The energy
values are:~$E=0.8$ (Figs. a and b),~$E=0.67$ (Figs. c and d),~$E=0.64$
(Fig. e).

Note that for small (close to the Thompson configuration), as well as for
sufficiently large, energies, the system's phase portrait is almost
regular (Fig.~\ref{pic4.eps} a, b, e). While for intermediate energy
values, the system's phase flow becomes almost completely chaotic
(Fig.~\ref{pic4.eps} c, d).

This work was supported in part by CRDF (RU-M1-2583-MO-04), INTAS (04-80-7297), RFBR
(04-205-264367) and NSh (136.2003.1).


\begin{thebibliography}{50}\itemsep=0pt

\bibitem{bmk-1}
A.~V.~Borisov, I.~S.~Mamaev,
{\scshape  Mathematical methods of vortex dynamics,} in ``Fundamental
and Applied Problems in the Theory of Vortices,"
A. V. Borisov, I. S.Mamaev, M.A. Sokolovskiy (Eds.), Moscow-Izhevsk: ICS 2003, 17--178  (in Russian).

\bibitem{bmk-2}
P.~ K.~Newton,
``The $N$-vortex problem. Analytical Techniques,"
Springer 2001.

\bibitem{bmk-3}
C.~L.~Charlier,
``Die Mechanik des Himmels,"
Berlin: W. de Gruyter \& Co. 1927.

\bibitem{bmk-4}
K.~M.~Khanin,
{\scshape Quasi-periodic motions of vortex systems,} Physica D, 4 (1982), 261--269.

\bibitem{bmk-6}
V.~V.~Kozlov,
``Symmetry, Topology and Resonances in Hamiltonian Mechanics," Springer 1996.

\bibitem{bmk-7}
S.~L.~Ziglin,
{\scshape Nonintegrability of the problem on motion of four point
vortices,} Dokl. Akad. Nauk SSSR, 250 (1990), no 6, 1296--1300 (in Russian).

\bibitem{bmk-8}
C.~C.~Lim,
{\scshape Graph theory and special class of symplectic transformations: the
generalized Jacobi variables,} J. Math. Phys., 1991 (32), 1--7.

\bibitem{bmk-9}
A.~V.~Bolsinov, A.~V.~Borisov, I.~S.~Mamaev,
{\scshape Lie algebras in vortex dynamics
and celestial mechanics --- IV,}
 Reg. \& Chaot. Dyn., 1999 (4), no 1, 23--50.

\bibitem{bmk-10}
A.~V.~Borisov, I.~S.~Mamaev,
``Poisson Structures and Lie Algebras in Hamiltonian
Mechanics," Izhevsk: Publ. House of Udm. Univer. 1999 (in Russian).

\bibitem{Limm}
C.~C.~Lim,
{\scshape A combinatorical perturbation method and Arnold's wiskered tori
in vortex dynamics,} Physica D, 1993 (64), 163--184.

\bibitem{Bagrets}
A.~A.~Bagrets, D.~A.~Bagrets,
{\scshape Nonintegrability of two problems in vortex
dynamics,} Chaos, 1997 (7), 368--375.

\bibitem{Aref1}
H.~Aref, N.~Pomphrey,
{\scshape Integrable and chaotic motions of four vortices. I. The case
of identical vortices,} Proc. R. Soc. London. 1982 (380 A), 359--387.

\bibitem{Eck1}
B.~Eckhardt, H.~Aref,
{\scshape Integrable and chaotic motion of four vortices.
II. Collision dynamics of vortex pairs,} Phil. Trans. R. Soc. Lond. 1988 (326 A), 655--696.

\bibitem{Bor_Leb1}
A.~V.~Borisov, V.~G.~Lebedev,
{\scshape Dynamics of three vortices on a
plane and a sphere --- II. General compact case,} Reg. \& Chaot. Dyn., 1998 (3), no 2, 99--114.

\bibitem{Bor_Leb2}
A.~V.~Borisov, V.~G.~Lebedev,
{\scshape Ddynamics of three vortices on a
plane and a sphere --- III. Noncompact case. Problems of collaps and
scattering,} Reg. \& Chaot. Dyn., 1998 (3), no 4, 76--90.

\bibitem{Bog2}
V.~A.~Bogomolov,
{\scshape Dynamics of vorticity at a sphere,} Fluid Dynamics, 1977
(6), 863--870.

\bibitem{82}
V.~V.~Kozlov,
{\scshape Integrability and nonintegrability in Hamiltonian
mechanics,} Russian Math. Surveys, 1983 (38), 1--76.

\bibitem{Eck}
B.~Eckhardt,
{\scshape Integrable four vortex motion,} Phys. Fluids. 1988 (31), 2796--2801.

\bibitem{95}
R.~Kidambi, P.~K.~Newton,
{\scshape Motion of three point vortices on a
sphere,} Physica D, 1998 (116), 143--175.

\bibitem{Uinteker}
A.~Wintner, ``The Analytical Foundations of Celestial Mechanics,"
Princeton
University Press 1941.

\bibitem{Cel}
A.~Celletti, C.~Falclini,
{\scshape A remark on the KAM theorem applied to
a four\1vortex problem,} J. Stat. Phys., 1988 (52), 471--477.


\end{thebibliography}
\end{document}